\documentclass[doublecol]{epl2} 

\usepackage{graphicx}
\usepackage{dcolumn}
\usepackage{amssymb}
\usepackage{bm}
\usepackage{epsfig}
\usepackage{psfrag}
\usepackage{setspace}

\def\3dots{\:\raisebox{-0.5ex}{$\stackrel{\textstyle.}{:}$}\:}
\def\beq{\begin{equation}}
\def\eeq{\end{equation}}
\def\bea{\begin{eqnarray}}
\def\eea{\end{eqnarray}}

\title{Irreversibility to Reversibility Crossover in Transient Response of an Optically Trapped Particle}
\shorttitle{Irreversibility to Reversibility Crossover}

\author{Manas Khan and A. K. Sood}
\shortauthor{M. Khan \etal}
\institute{Department of Physics, Indian Institute of Science, Bangalore - 560012, India                    
}


\abstract{We study the transient response of a colloidal bead which is released from different heights and allowed to relax in the potential well of an optical trap. Depending on the initial potential energy, the system's time evolution shows dramatically different behaviors. Starting from the short-time reversible to long-time irreversible transition, a stationary reversible state with zero net dissipation can be achieved as the release point energy is decreased. If the system starts with even lower energy, it progressively extracts useful work from thermal noise and exhibits an anomalous irreversibility. In addition, we have verified the Transient Fluctuation Theorem and the Integrated Transient Fluctuation Theorem even for the non-ergodic descriptions of our system.}

\begin{document}

\maketitle
%\begin{multicols}{2}

%\narrowtext

Since the origin of the second law of thermodynamics which defined a direction of time, a question is often asked: how the irreversible property emerges from the reversible dynamics at microscopic scales \cite{debate}. A detailed examination of the problem reveals that the mystery lies in its spontaneous energy dispersal as dissipation. It makes us curious about the time evolution of small stochastic systems where the thermal fluctuations become comparable to the dissipative loss. It is only in the last decade that this interesting issue has been addressed in terms of the Fluctuation Theorems \cite{evanFT, cohenFT, jarzynskiFT, crookFT}. Evans and Searles Fluctuation Theorem (FT) \cite{evanFT} gave a quantitative description of irreversibility, providing an analytical expression for the probability of the system trajectories that are associated with a negative dissipative flux. As expected, those `anti-trajectories' become visible only for short time scales in microscopic systems. Laboratory experiments with bio-molecules manipulated mechanically \cite{science, nature} or colloidal particles in optical trap \cite{evanDrag, evanCapture, pnas, bechinger, cilibertoPRL, cilibertoEPL, evanVisco} have been reported subsequently corroborating the theorem(s).

In this Letter we address this important issue of reversibility with a closer focus on its main controlling factor, the initial potential energy of the system. We show experimentally how a system's initial energy in comparison to the background thermal noise, entirely dictates its time evolution. Starting with very low energy, the system can even continue to convert the heat fluctuations into useful work as it evolves in time. In particular, we have studied the transient response and the reversible behavior of a colloidal bead in a harmonic potential well - created by a very weak optical trap. As the bead is released from different points in the potential well landscape with varying initial energy and allowed to equilibrate, the time evolution and the reversibility of the system show a dramatic change in their behavior. If particle starts with a potential energy higher than the average thermal energy of the bath, its mean motion is downhill but the underlying reversibility of the dynamics generates occasional uphill motions. However, with the release point going down the potential well, the reversibility i.e. the probability of performing along the backward direction increases accompanied by a diminished dissipation. At some point, it becomes completely reversible when the forward and backward motions occur with equal probability. If the bead is released from a point which is further down the potential well, the heat flux starts flowing along the opposite direction prevailing the dissipative loss and pushes the system to a higher energy state making the system irreversible again, in the converse direction. These exceptional energy-gaining motions are ultimately a consequence of the micro-reversibility of the system. The results presented in this Letter demonstrate experimentally the competition between micro-reversibility and irreversible dissipation in governing the long-time reversibility of a system.

In our study, we have evaluated the dissipation function, $\Omega_{t}$, as a measure of reversibility \cite{evanPRE}. Let $P(\textbf{r}_{0} , \textbf{r}_{t})$ be the probability of finding a set of system trajectories $\left\{\textbf{r}_{0} , \textbf{r}_{t}\right\}$ (forward trajectories) evolving spontaneously from $\textbf{r}_{0}$ to $\textbf{r}_{t}$ over a time $t$, dissipating positive energy to the surrounding. $P(\textbf{r}_{t} , \textbf{r}_{0})$ represents the probability of occurrence of the corresponding backward trajectories or `anti-trajectories' $\left\{\textbf{r}_{t} , \textbf{r}_{0}\right\}$ of the same time duration $t$, where the thermal noise does useful work and the system gains energy. The measure of reversibility, $\Omega_{t}$, is defined as $\Omega_{t} (\textbf{r}_{0} , \textbf{r}_{t}) = ln\left[P(\textbf{r}_{0} , \textbf{r}_{t})/P(\textbf{r}_{t} , \textbf{r}_{0})\right]$, where $\Omega_{t} = 0$ corresponds to perfect reversibility. For a thermostatted dissipative system which is allowed to relax from an initial equilibrium state, the distribution and the time evolution of the dissipation function, $\Omega_{t}$, follows two forms of FTs \cite{evanCapture, evanPRE} - the Transient Fluctuation Theorem (TFT):  
\begin{equation}
\frac{P\left( \Omega_{t} = A \right)}{P\left( \Omega_{t} = - A \right)} = exp(A)
\end{equation}
and the integrated form, namely the Integrated Transient Fluctuation Theorem (ITFT):
\begin{equation}
\frac{P\left( \Omega_{t} < 0 \right)}{P\left( \Omega_{t} > 0 \right)} = \left\langle exp(- \Omega_{t})\right\rangle_{\Omega_{t} > 0}
\end{equation}
In this work, we show the distribution and the time evolution of the dissipation function, $\Omega_{t}$, as a quantitative description of the transient response of the released bead. Until the release at $t=0$, the bead is held in equilibrium (by an additional very strong optical trap) at the release point. Depending on the release point energy, the time responses of $\Omega_{t}$ show three distinct possible behaviors. Either it becomes more positive with time as expected, or anomalously decays down to the negative side for a very low energy release thereby undergoing a reversible-to-irreversible transition in both the cases although along the opposite directions. In an intermediate case, the system continues to remain reversible.

%FIGURE-1
\begin{figure}[htb]
  % Requires \usepackage{graphicx}
%\includegraphics[width=0.65\textwidth]{Fig1.eps}\\
\includegraphics[width=0.43\textwidth]{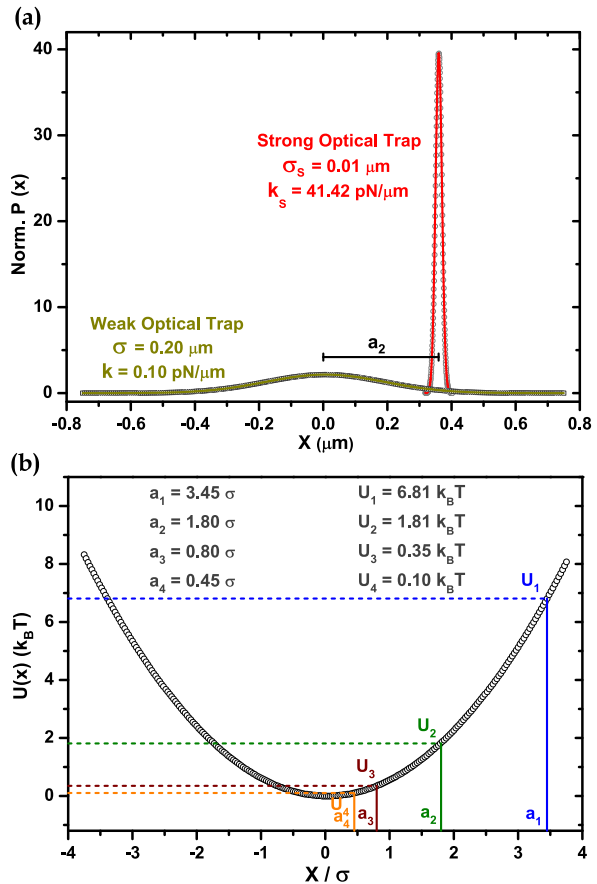}\\
\caption{(Color online) (a) The normalized position distribution of a trapped bead for both the traps are shown for comparison. (b) The potential well created by the weak optical trap. The release points $a_{i} (i=1:4)$ and the corresponding energies $U_{i} (i=1:4)$ are shown on the potential well graph. The potential well is constructed from the normalized position distribution of a trapped bead (shown in (a)), following $U(x) = -ln[P(x)]$ and $U(0)=0$. } \label{potential}
\end{figure}

Using a dual optical trap setup, a silica bead has been released from a set of points $a_{i} (i = 1:4)$, with corresponding potential energies $U_{i} (i=1:4)$, in a harmonic potential well landscape (Fig. \ref{potential}b). An infrared (1064 $nm$) laser beam has been split into two polarization components by using a $\lambda/2$ plate and a polarizing beam splitter to create two independently controllable traps at the focal plane of a $100\times$, $1.4 NA$ objective fitted to an inverted microscope. The ratio of the two polarizations was set to an extreme value to make one of them a very weak trap (stiffness: $k$ = 0.1 $pN/\mu m$, standard deviation (SD) = $\sigma$). A very dilute aqueous suspension of 1.61 $\mu m$ diameter ($d$) silica beads, loaded in a sample cell with 125 $\mu m$ spacer, was used for all of the experiments. A high speed CCD camera attached to the microscope captured the bead's dynamics in the trap and then a Matlab based particle tracking program \cite{PPT} was utilized to track the center of the bead from the recorded image sequences. The normalized position distribution of a trapped bead, (Fig. \ref{potential}a) provides the stiffness of the trap (along $x$) as $k = k_{B}T/\left\langle x^{2}\right\rangle = k_{B}T/\sigma^{2}$. To release a colloidal bead from different points, the second trap (very strong; stiffness: $k_{s}$ = 41.42 $pN/\mu m$, SD: $\sigma_{s} = \sigma/20$), movable along the $x$ axis, was utilized. The values of the relaxation time of the bead in the traps, given by $\tau = 3 \pi \eta d/k$ where $\eta$ is the viscosity of water, are 0.15$s$ and 0.37$ms$ for the weak and the strong trap respectively. Both the traps were placed precisely at the same $Z$ plane, well above ($\sim$ 20 $\mu m$) the bottom plate. The strong trap was centered at the release point ($x = a$, $y = 0$) (Fig. \ref{potential}a) to hold the particle in equilibrium until the beam is chopped to release the bead. As the bead was released at $t = 0$, from $x = a_{i}$($i=1:4$) (potential energy $U_{i}$), its transient trajectories were recorded at 500 frames per second. To re-initiate the trajectories automatically with the same initial phase point, the strong trap-beam was chopped at 3 $Hz$. Only those trajectories were considered as valid trajectories where the distance between the trap centers (i.e. the value of $a$) was not changed by more than 10\% from the set value because of slight drift of the laser beams during the experiment. Over 3000 such trajectories were captured for each of the release points.

The equipartition method which has been used in this study to calculate the trap stiffness gives accurate values only when the viscous relaxation time of the bead in the trap is much longer than the exposure time (0.5 $ms$ for our experiments) of the camera \cite{blur}. In case of the weak trap potential this condition is well satisfied and there is no motion blur to affect the value of $k$. Though there would be a small correction factor ($\sim$ 7\%) for $k_{s}$ because of the motion blur, it is not a matter of concern as all our results are independent of the accurate evaluation of the strong trap stiffness.

\begin{figure}[htb]
  % Requires \usepackage{graphicx}
\includegraphics[width=0.45\textwidth]{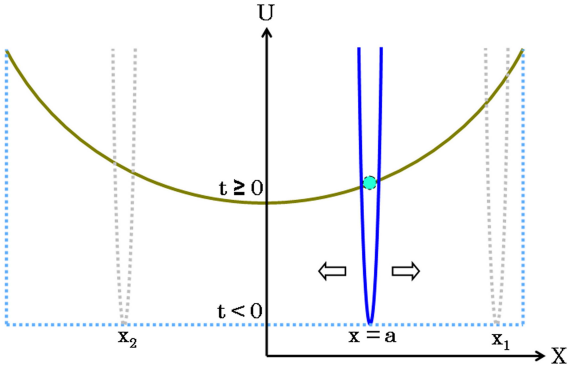}\\
\caption{(Color online) A movable strong trap can perform a scanning motion along the $X$ axis and release the particle from each and every $x$ value with equal probability. This protocol is statistically equivalent to the protocol where the initial ($t < 0$) potential is flat which is changed to the weak harmonic potential at $t = 0$. In both the cases the particle could be released from any initial $x$ with equal probability.} \label{model}
\end{figure}

It would be important to mention that this experiment, for theoretical considerations, can be treated as a practical realization of the scenario where an absolutely flat potential (a harmonic potential well with stiffness $k_{0}\rightarrow 0$) at $t < 0$ is abruptly changed to a harmonic potential well with stiffness $k$ at $t = 0$, as depicted in Fig. \ref{model}. In the flat potential the particle resides at any $x$ with equal probability (at $t < 0$) and as it is changed to the weak trap potential at $t = 0$, the starting point of the particle's trajectory ($r_{0}$) takes any $x$ value with equal likeliness. However, our aim in this experimental study is to investigate the time evolution of only those trajectories that start from some specific initial points, $x = a_{i}$ ($i=1:4$). Toward a practical realization of this protocol, the flat potential can effectively be created with a movable strong trap. One way of doing it would be making the strong trap scan along the $X$-axis at a very high speed so that the colloidal bead feels an averaged out flat potential. It can also be achieved in a more organized way, as has been done in our experiment. A strong trap (stiffness $k_{s}$) can be placed at a point, e.g. $x_{1}$ as shown in Fig. \ref{model}, holding the particle (at $t<0$) until it is released to the weak potential (stiffness $k$) at $t=0$ to construct a system trajectory with $r_{0} = x_{1}$. The strong trap is then moved to a neighboring point $(x_{1} - \delta x)$ at the left side in a repeat experiment (using a nano-tilt mirror mount, a $\delta x$ of as small as 5 $nm$ can be achieved in our setup) to release the particle from $r_{0} = (x_{1} - \delta x)$. This process can be repeated a large number of times so that the strong trap scans through the $X$-axis several thousand times. In this way system trajectories with $r_{0}$ having all possible $x$ values with equal probabilities and their corresponding $r_{t}(t)$ can be acquired. Since the focus of our study is on the system trajectories with the starting points ($r_{0}$) being \{$x = a$, $y = 0$\} ($a = a_{i}$; $i=1:4$), we have discussed here only those cases when the strong trap is positioned at the four respective points.

The time-dependent probability of a colloidal bead trapped in a harmonic potential is known analytically by solving the Langevin equation \cite{RMP, probability}. Following this, we can construct the probability distribution $P(\textbf{r}_{0} , \textbf{r}_{t})$ and in turn the dissipation function, $\Omega_{t}$, in terms of our measured variables. Knowing a bead's initial position $\textbf{r}_{0}$ in an optical trap of stiffness $k$, the probability of finding it at $\textbf{r}_{t}$ after time $t$ can be given by \cite{RMP, probability}, 
\begin{eqnarray}
P(\textbf{r}_{t} ; \textbf{r}_{0}, k, t) =  \left(\frac{k}{2 \pi k_{B} T [1-exp(-2t/\tau)]} \right) \nonumber \\
 \times exp \left(- \frac{k[\textbf{r}_{t} - \textbf{r}_{0} exp(-t/\tau)]^{2}}{2k_{B}T[1-exp(-2t/\tau)]}\right), 
\end{eqnarray}  
where $\tau = 3 \pi \eta d /k$ is the characteristic relaxation time in the trap potential, $\eta$ being the viscosity of the medium. Just before the release, the colloidal bead resides in the strong trap (centered at $x = a$) in equilibrium, and, therefore, the position distribution of the bead in the strong trap can be given by the time-independent Boltzmann distribution, $P_{B}(\textbf{x}, k_{s})$ \cite{evanPRE, probability}. As the stiffness value of the strong trap is very large compared to the other parameters, the distribution can practically be treated as a delta function \cite{support} at $x = a$. Therefore, $P(\textbf{r}_{0} , \textbf{r}_{t})$ becomes equal to $P(\textbf{x}_{t} ; \textbf{a}, k, t)$. Similarly, the probability of the corresponding backward trajectories would be $P(\textbf{r}_{t} , \textbf{r}_{0}) = P(\textbf{a} ; \textbf{x}_{t}, k, t)$. With these two probabilities in hand, the analytical form for the dissipation function becomes:
\begin{eqnarray}
\Omega_{t} (\textbf{a},\textbf{x}_{t}) = \frac{k}{2 k_{B} T} \left( a^{2} - x^{2}_{t}\right) 
\end{eqnarray}  
Though the explicit time dependence, as seen in Eq. 3, gets canceled in this expression, $\Omega_{t}$ has a strong time dependence through $x_{t}$, which is the instantaneous position of the particle at time $t$. It would be pertinent to note that the dissipation function as defined above, is actually the energy dissipated; in general the heat lost ($-\Delta Q$) over a trajectory (initiated at ${x}_{0}={a}$) of time duration $t$ divided by the reservoir temperature.

% Requires \usepackage{graphicx}
\begin{figure*}[htbp]
\includegraphics[width=1\textwidth]{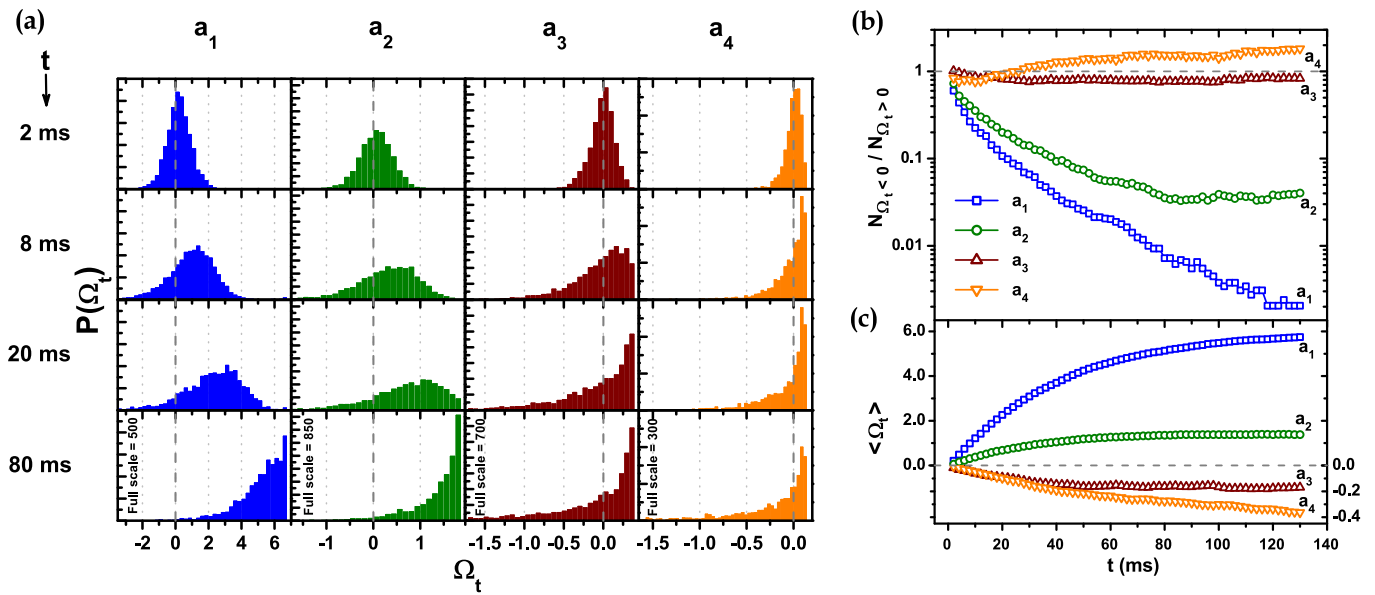}\\
\caption{(Color online) (a) The distributions of the dissipation function, $P\left(\Omega_{t}\right)$, are shown for all the release points $a_{i}(i=1:4)$ at four subsequent time slices. The horizontal and vertical scales are same for all the four panels in a column, as indicated at the bottom of each column. (b) The ratios of number of backward trajectories ($N_{\Omega_{t}<0}$) to the number of forward trajectories ($N_{\Omega_{t}>0}$) have been plotted with $t$ for different release points. The dashed horizontal line indicates ideal reversibility (c) The time evolutions of the ensemble averages of $\Omega_{t}$ are plotted. The negative side is shown in a magnified scale (at the right side) for better clarity. The legends are same as those in (b). Here too, the horizontal dashed line corresponds to ideal reversible behavior.} \label{df}
\end{figure*}

The dissipation function, $\Omega_{t}$, for this study does not depend on the strong trap stiffness ($k_{s}$) unlike the cases of some other seemingly similar experiments \cite{evanCapture, evanVisco}. Our goal here is to investigate the transient trajectories of a colloidal particle subjected to the restoring force of a weak optical trap (stiffness $k$). The weak optical trap sets up a potential energy landscape that perturbs the trajectories of an otherwise free particle. The system's response to that perturbation, more specifically the reversibility in its time evolution has been quantified through the dissipation function here. The expression for $\Omega_{t}$ , as defined in Eq. 4, is complete and unambiguous for this purpose. The strong trap is utilized only to re-initiate the particle trajectories from a predefined specific point (${x}_{0}={a}$ at $t=0$) in the weak trap potential landscape. This mechanism enables us to record a large number of system trajectories characterized by the definite initial potential energies ($U$, corresponding to each $a$) and to study their time evolutions. In contrast, the dissipation functions as defined in the other reported experiments \cite{evanCapture, evanVisco} describe the system response to an external perturbation that is created by a small discontinuous change of the trap stiffness and therefore comprise of both the trap stiffnesses, before and after the change.

We have analyzed the dissipation function, $\Omega_{t}$, for four subsets of the system trajectories characterized by their starting points ($x_{0}=a_{i}$; $i=1:4$), which uniquely define the initial potential energies. The distributions of $\Omega_{t}$ for these four groups of trajectories are  shown in Fig. \ref{df}a. Irrespective of the release points, the distributions at very short time (2 $ms$) are symmetric about zero indicating the reversible nature of the system at this time scale. As time increases, the distributions no longer remain symmetric. For release points $a_{1}$ and $a_{2}$, not only the peaks but the whole distributions shift to the positive side with time. In these cases, where the system starts with a higher energy, the time evolutions are overwhelmingly dominated by the trajectories along which the system dissipates its energy and eventually goes to a lower energy state - sliding down the potential well irreversibly. For the trajectories that originate at $a_{3}$, the distribution of the dissipation function does not evolve much after a very short transient time (20 $ms$). Though the distribution at a later time is peaked at a small positive value, a compensating long tail in the negative side nearly equals the probability of $\Omega_{t}$ being positive or negative. The system, in this case, starts with a low energy state and performs reversibly as the forward and backward trajectories become equally probable. When the bead is released from $a_{4}$, with potential energy very close to its minimum value, the peak at the positive side decreases with time and the negative side becomes increasingly populated with the tail growing longer. Here the backward trajectories prevail as the system continues to convert the heat fluctuations into useful work progressively and thereby climbs to higher energy states - making the time evolution irreversible again, along the opposite direction.

For a better quantitative description of the system's time evolution, the two important quantities - the relative probability of observing backward trajectories ($\Omega_{t}<0$) in comparison to the forward trajectories ($\Omega_{t}>0$), given by the ratio: $N_{\Omega_{t}<0}/N_{\Omega_{t}>0}$ (Fig. \ref{df}b), and the ensemble averages of the dissipation function, $\left\langle \Omega_{t}\right\rangle$ (Fig. \ref{df}c), are plotted against time, $t$. While the relative probability is a direct indication of the system's reversibility, the other quantity, $\left\langle \Omega_{t}\right\rangle$, stands for the average dissipative loss of the system. As the system starts with varying potential energies, the time evolution of these two quantities change dramatically. For higher initial energies ($a_{1}$ and $a_{2}$), the system behaves reversibly only for a very short time and then turns completely irreversible with a large dissipative loss. With decreasing initial energy ($a_{1}$ to $a_{3}$) the system becomes increasingly reversible as the heat loss due to dissipation diminishes. Finally, starting from $a_{3}$, the system exhibits perfect reversibility, accompanied by a very small intake of heat from the bath. If the system's initial energy goes even lower (at $a_{4}$), the backward trajectories start dominating with time making the system irreversible again. In this case, the system extracts useful work from the thermal fluctuations and performs increasingly along the reverse direction.

It is noteworthy that the system's time evolution undergoes an interesting transition at $a_{3}$. When the system starts from a point which is above $a_{3}$, it evolves predominantly along the forward trajectories which signify minimization of energy, along with spontaneous production of entropy. As the starting point goes below $a_{3}$, the system can no longer minimize its energy simultaneously with generation of entropy. Therefore, the system performs increasingly along the backward trajectories to maximize the entropy at the cost of gaining potential energy. This demonstrates a very fundamental and intriguing fact of statistical mechanics. Starting from an initial phase point, an isolated system always tries to expand its probability distribution by exploring the phase space accessible to it and thereby producing entropy. Usually, in this process, the system goes to lower energy phase points and dissipate energy to the surroundings. However, if a system starts from near the bottom of a potential landscape, most of the accessible phase points are at higher energies and the system explores those higher energy phase points to maximize its entropy. In this process of spreading its probability density, the system extracts energy from the heat bath and performs positive work on an external system to go to the higher energy states.

It is also worth mentioning that here the long-time-average of $\Omega_{t} (x_{0} = a_{i}; i=1:4)$ does not comply with the average entropy production, as we have considered only the specific subsets of system trajectories to evaluate $\Omega_{t}$. If all the system trajectories starting from all possible phase points are taken into account, the ensemble average, $<\Omega_{t}>$, would always be positive, in accordance with the average entropy production.

\begin{figure}[htb]
% Requires \usepackage{graphicx}
%\includegraphics[width=0.7\textwidth]{Fig3.eps}\\
\includegraphics[width=0.4\textwidth]{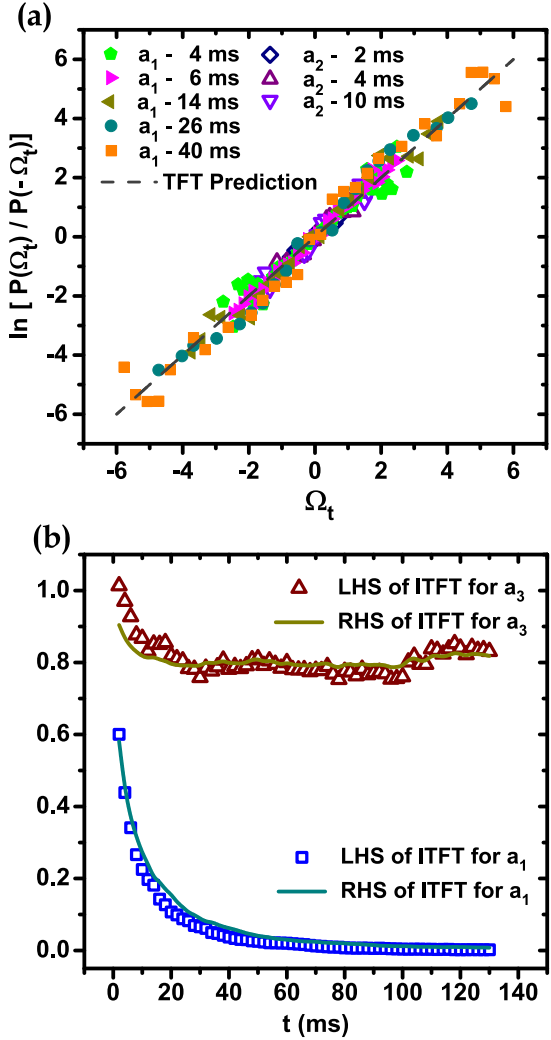}\\
\caption{(Color online) (a) Logarithm of the ratio of the probability to find trajectories with $\Omega_{t}=A$ to those with $\Omega_{t}=-A$ has been plotted against the value $\Omega_{t}=A$, for a few different cases. The end points deviates a bit from the prediction of TFT due to the poor statistics at the tail of the distribution. (b) The LHS of the ITFT (same as Fig. \ref{df}b) has been superimposed on the RHS of the ITFT, $\left\langle exp(-\Omega_{t})\right\rangle_{\Omega_{t}>0}$, for release points $a_{1}$ and $a_{3}$.}
\label{ft}
\end{figure}

Even though the dissipation functions $\Omega_{t} (x_{0} = a_{i}; i=1:4)$ evaluated in our study do not comprise all the system trajectories, rater contain only a specific small subset (defined by $x_{0} = a$) at a time, and thereby fail to satisfy the requirement of being ergodic description of the system, nonetheless they satisfy the TFT and the ITFT separately for the subsets. To verify the TFT (Eq. 1), the logarithm of the ratio of the number of trajectories with a particular value of $\Omega_{t}$, say $A$, to those with the value of $-A$, has been plotted with the $\Omega_{t}$ value, i.e. $A$. Wherever the distribution of $\Omega_{t}$ has a wide spread about zero, the slope of the straight line fit to the plot takes a value within $1 \pm 0.06$, showing very good agreement with the TFT prediction. Some of those plots have been shown in Fig. \ref{ft}a, superimposing with the TFT dictated straight line of slope $1$. For the cases where the distributions are narrow or significantly skewed, the statistics are not sufficient to compare and comment conclusively on the agreement or disagreement of the experimental results with the TFT prediction. To verify the integrated form, the ITFT (Eq. 2), the LHS (same as Fig. \ref{df}b) has been compared with the RHS for the release points $a_{1}$ and $a_{3}$ in Fig. \ref{ft}b and as displayed, they match excellently. The verification of the TFT and the ITFT for these non-ergodic system descriptions is intriguing as by definition the theorems demand ergodicity i.e. the dissipation function needs to be calculated over all the possible system trajectories in order to satisfy the theorems.

In summary, we have investigated the transient response of a colloidal bead that is released from different energies in the potential well landscape created by an optical trap and is allowed to relax. In addition to the verification of TFT and ITFT even for non-ergodic descriptions of this system, we have shown that the reversibility of the bead's response is entirely determined by its initial potential energy. At short time, the system behaves reversibly irrespective of the release points. As time increases, the system undergoes an expected reversible-irreversible transition for a higher energy release. However, when released from a specific lower energy, the system remains perfectly reversible for exceptionally long time, with almost zero net dissipative flux. For a release with even lower initial free energy which is very close to its lowest value, the thermal fluctuations start providing useful work thereby forcing the system to perform increasingly along the reverse direction, which leads to an anomalistic irreversible state as the system evolves in time.

\acknowledgments

We thank Council for Scientific and Industrial Research (CSIR), India for financial support and Prof. S. Ramaswamy for useful discussions.


\begin{thebibliography}{0}
\bibitem{support}
\bibinfo{author}{To justify this, we have checked the distribution of the bead's displacement in the first time step (2 mS) after release and it is much wider ($\sigma > 4 \times \sigma_{s}$) compared to its position distribution in the strong optical trap.}
\end{thebibliography}
\end{document}